\documentclass[fleqn,twoside]{article}%
\usepackage{amsmath}
\usepackage{amsfonts}
\usepackage{amssymb}
\usepackage{graphicx}
\usepackage{cite}
\def\be{\begin{equation}}
\def\ee{\end{equation}}
\def\bi{\bibitem}
\begin{document}
\title{Self-Similar Magnetohydrodynamics.}
\author{Abhik Kumar Sanyal$^1$ and D. Ray $^2$}
\maketitle
\noindent
\begin{center}
\noindent
$^1$ Dept of Physics, University College of Science,\\
92 A.P.C. Road, Calcutta-700009, India.\\
$^2$ Dept of Applied Mathematics, University College of Science,\\
92 A.P.C. Road, Calcutta-700009, India.\\
\end{center}
\footnotetext{\noindent
Electronic address:\\
\noindent
$^1$ sanyal\_ ak@yahoo.com \\
Present address: Dept. of Physics, Jangipur College, India - 742213.}
\noindent
\abstract{For the solution of the full set of magnetohydrodynamics (MHD) equations in the presence of gravity due to a central point-mass, a  self-similar theory for a general polytrope has already suggested a set of exact time-dependent solutions by analytical methods for a $\gamma =  {4\over3}$ polytrope, since $\gamma = {4\over3}$ is the simplest to treat. In the present paper while going for a complete set of self-similar  solutions, we find that $\gamma = {4\over3}$ is the only physically-possible polytrope and that then, pressure is independent of the scalar function $A$ and depends on angle and time only. We also obtain a specific form of time-dependence for the self-similar variable.}\\
\maketitle
\flushbottom
\section{Introduction:}
So far, attempts made to account for the presence of gravity in many magnetohydrodynamic processes have given rise to some mathematical problems, which are difficult to treat. These problems have been tackled by considering the magnetic field in a static equilibrium with a statified atmosphere by Dungey \cite{D} and Low \cite{L} and recently by Tsinganos \cite{T}. In the present paper we consider time-dependent
magnetohydrodynamic (MHD) equations in the presence of gravity due to a central point-mass, which is useful in describing stellar structure.\\
Now a magnetic field model of enough interest must involve at least two spatial variables, since a one-dimensional field give rise to some properties of MHD which are precluded. Time-dependent problems involving two spatial variables are difficult to handle. So the idea of self-similar theory is  applied here. Self-similar solutions are the leading terms in an asymptotic expansion of a non-self-similar evolution, far from the initial conditions and far from the influence of boundary conditions. \\
Some self-similar solutions of time-dependent MHD were first given by Bernstein and Kulsrud \cite{BK} and Kulsrud et al. \cite{Kea}. But both of them neglected the effect of gravity. Self-similar theory has been successfully applied ir, magnetohydrodynamics in the presence of gravity due to a central point-mass by Low \cite{L2}. In the present paper we shall follow the procedure given by Low and as such extend his work, giving some important physical solutions.\\
The complete set of ideal MHD equations is:
\be \rho {\partial \vec v\over\partial t} + \rho(\vec v . \nabla)\vec v = {1\over 2\pi} (\nabla \times \vec B)\times \vec B - \nabla p - \rho{GM\over r^2}\hat r,\ee
\be \nabla .\vec B = 0, \ee
\be {\partial \vec B\over \partial t} = \nabla \times (\vec v \times \vec B),\ee
\be {\partial \rho\over \partial t} + \nabla . (\rho .\vec v) = 0, \ee
\be {\partial\over \partial t}[\log{p\rho^{-\gamma}}] + \vec v . \nabla [\log{p\rho^{-\gamma}}] = 0.\ee
Where, $B, ~v, ~p, ~p$ are the magnetic field, velocity field, density and pressure respectively. Gravity due to a spherically symmetric-star of mass $M$ is included by Low \cite{L2}, where $G$ is the Newtonian gravitational constant and $r$ is the distance from the centre of the star. MHD flow is considered to be isentropic with polytropic index $\gamma$. Low \cite{L2} attempted to find self-similar solutions (which represent the intrinsic evolution behaviour of a system not dependent on the incidental details of particular initial or boundary conditions) of MHD such that the time evolution is controlled by the self-similar radial variable,
\be\zeta = {r\over \Phi} \ee
where $\Phi$ is a function of time. Now using polar coordinates $r, \theta, \phi$ and axisymmetric magnetic field can be expressed in terms of two scalar functions $A$ and $B$ \cite{SC},
\be \vec B = {1\over r \sin{\theta}}\left( {1\over r}{\partial A \over \partial \theta}\hat r + {\partial A \over \partial r}\hat\theta + B \hat\phi \right)\ee
Assuming a global fluid-flow strictly in the radial direction $\vec v = v(r, \theta, t)\hat r$, equations (1) - (5) take the following forms:
\be \rho\left({\partial v\over \partial t} + v {\partial v\over \partial r}\right) = {1\over 4\pi r^2 \sin^2{\theta}}\left(LA{\partial A\over \partial r} + B {\partial B\over \partial r}\right) -{\partial p\over \partial r} - {GM\rho\over r^2}.\ee
\be {1\over 4\pi r^2 \sin^2{\theta}}\left(LA{1\over r}{\partial A\over \partial \theta} + B {1\over r}{\partial B\over \partial \theta}\right) + {1\over r}{\partial p\over \partial \theta} = 0.\ee
\be {1\over r}{\partial (B, A)\over \partial (r, \theta)} = 0.\ee
\be {\partial A\over \partial t} + v {\partial A\over \partial r} = 0.\ee
\be {\partial B\over \partial t} + {\partial\over \partial r}(v B) = 0. \ee
\be {\partial \rho\over \partial t} + {1\over r^2} {\partial \over \partial r}(r^2 \rho v) = 0.\ee
\be {\partial\over \partial t}[\log{p\rho^{-\gamma}}] + v {\partial\over \partial r} [\log{p\rho^{-\gamma}}] = 0,\ee

where, \be L = {\partial^2\over \partial r^2} + {\sin{\theta}\over r^2}{\partial\over \partial \theta}\left({1\over \sin{\theta}}{\partial\over \partial \theta}\right), \ee
and, $A = A(r,\theta, t)$, $B = B(r,\theta, t)$, $\rho = \rho(r,\theta, t)$, and $p = p(r,\theta, t)$. Now in order to study self-similar solutions of MHD critically, we have changed the variables in the following section 2 and then tried to find out the solutions of the MHD equations by applying the self-similar-condition (6) as suggested by Low \cite{L2}, in Section 3, which ultimately give rise to a fairly interesting result.
\section{Change of Variables:}
Equation (10) implies,
\be B = B(A, t).\ee
We now consider equation (12), make the change of variables in the form $B = {\partial C\over\partial r}$, and $-v B = {\partial C\over \partial t}$, so that we arrive at, ${\partial C\over \partial t} + v {\partial C\over \partial r} = 0.$
Comparing the above equation with equation (11) we get,
\be C = C(A, \theta).\ee
Since $r$ and $t$ are independent variables. So equation (13) can be written in the following form,
\be {\partial \over \partial t}(r^2 \rho) + {\partial \over \partial r} (r^2\rho v) = 0.\ee
Hence, there must exist a function $D$ such that
\be {\partial D\over \partial r} = r^2 \rho,~~~~~\mathrm{and}, ~~~{\partial D\over \partial t} = -r^2 \rho v.\ee
and hence
\be {\partial D\over \partial t} + v{\partial D\over \partial r} = 0.\ee
comparing this with equation (11) we get,
\be D = D(A, \theta).\ee
Finally comparing equations (11) and (14) we get,
\be p\rho^{-\gamma} = E(A, \theta).\ee
Now to simplify the calculations involved in this paper, we change the set of
independent variables from $(r, \theta, t )$ to $(A, \theta, t)$. So that $v = v(A, \theta, t),~~ r = r(A, \theta, t ),~~p = p(A, \theta, t )$ and so on. From equation (11) we get,
\be v = -{A_t\over A_r} = {\partial r\over \partial t} = r_t.\ee
Also,
\be B = {\partial C\over \partial r}|_{\theta, t} = C_A A_r = {C_A\over r_A}.\ee
\be r^2\rho = {\partial D\over \partial r}|_{\theta, t} = D_A A_r = {D_A\over r_A}.\ee
Therefore,
\be \rho = {D_A\over r^2 r_A}.\ee
and from (21)
\be p = E\left[{D_A\over r^2 r_A}\right]^\gamma.\ee
The dependent-variables are connected with the old $(r, \theta, t)$ and the new $(A, \theta, t)$ set of coordinates in the following manner:

\be \begin{split} & v_{told} = v_{tnew} - v_r r_t,\\&
p_{rold} = p_A r_A,~~~~~p_{\theta old} = p_\theta - p_A {r_\theta\over r_A},\\&
A_{rold} = {1\over r_A};~~~~~A_{\theta old} = {r_\theta\over r_A}. \end{split}\ee
So with these new set of variables we can now move on to study the self-similar solutions of MHD as follows.

\section{Study of the Solutions:}

All calculations done in this section are in the set of coordinates $(A, \theta, t)$ until the boundary condition (6) for self-similar solutions is being applied.

\be \left[\rho\left(v_t + v v_r + {GM\over r^2}\right) + p_r\right] A_{\theta} = p_\theta A_r.\ee
Using equation (23) in the above equation, we get,

\be \left[\rho\left(v_t + r_t v_r + {GM\over r^2}\right) + p_r\right] A_{\theta} = p_\theta A_r.\ee
Now for the coordinate transformation from $(r, \theta, t)$ to $(A, \theta, t)$, we apply equations (28) in the above equation and obtain,

\be \left[\rho v_t + {p_A\over r_A} + {GM\rho\over r^2}\right] \left[-{r_\theta\over r_A}\right] = \left[p_\theta - p_A{r_\theta\over r_A}\right]\left[{1\over r_A}\right],\ee
which may be simplified to

\be  \rho\left( v_t + {GM\over r^2}\right) r_\theta = - p_\theta\ee
Again using equation (23), (26) and (27) in the above equation, we get,

\be \left(r_{tt} + {GM\over r^2}\right)\left({D_A r_\theta\over r^2 r_A}\right) = -\left[E\left({D_A\over r^2 r_A}\right)^{\gamma}\right]_\theta.\ee
Now, equation (9) can be written in the fo!lowing manner,

\be {1\over 4\pi r^2 \sin{\theta}}(L A A_\theta + B B_\theta) + p_\theta = 0,\ee
which may be reduced to

\be {1\over 4\pi r^2 \sin{\theta}}(L A  + B B_A)A_\theta + p_\theta = 0.\ee
Now operating L as given in equation (15) on $A$ the above equation takes the form:

\be {1\over 4\pi r^2 \sin^2\theta}\left(A_{rr} + {A_{\theta\theta}\over r^2} - {\cot\theta\over r^2}A_\theta + B B_A\right)A_\theta + p_\theta =0.\ee
Now from equation (28), we have

\be A_\theta = - {r_\theta\over r_A};~~~~~A_r =  {1\over r_A};~~~~~A_{\theta\theta} = \left({r_\theta\over r_A}\right)_A\left({r_\theta\over r_A}\right);~~~~~A_{rr}= \left({1\over r_A}\right)_A\left({1\over r_A}\right).\ee
Using these relations together with equation (24), equation (36) turns out to be,

\be {1\over 4\pi r^2 \sin^2{\theta}}\left[\left({1\over r_A}\right)_A {1\over r_A} + {1\over r^2}\left({r_\theta\over r_A}\right)_A + {\cot{\theta}\over r^2}{\left({r_\theta\over r_A}\right)} + \left({C_A\over r_A}\right)_A{C_A\over r_A}\right]\left(-{r_{\theta}\over r_A}\right) + p_\theta -p_A {r_\theta\over r_A} = 0.\ee
or,

\be \left[\left({1\over r_A}\right)_A  + {r_\theta\over r^2}\left({r_\theta\over r_A}\right)_A + \left({\cot{\theta}\over r^2}\right){r_\theta} + C_A\left({C_A\over r_A}\right)_A + p_A r_A\right]r_\theta = 4\pi r^2 r_A^2 \sin^2{\theta}~p_\theta.\ee
Now Using equation (27), we finally obtain,

\be \begin{split}&\left[\left({1\over r_A}\right)_A  + {r_\theta\over r^2}\left({r_\theta\over r_A}\right)_A + \left({\cot{\theta}\over r^2}\right){r_\theta} + C_A\left({C_A\over r_A}\right)_A + r_A \left\{E\left({D_A\over r^2r_A}\right)^\gamma\right\}_A\right]r_\theta\\&
 \hspace{3.0 in}= 4\pi r^2 r_A^2 \sin^2{\theta}\left\{E\left({D_A\over r^2r_A}\right)^\gamma\right\}_\theta.\end{split}\ee
As mentioned, we seek self-similar solutions of MHD equations following Low \cite{L2}, such that time evolution is controlled by the self-similar radial variabie,

\be \xi = {r\over \Phi}\ee
where, $\Phi$ is a function of time. According to Low \cite{L2}, let us assume that $A$ varies with time and radial distance not independently, but through the self-similar variable $\xi$ given by equation (41) for some function $\Phi$. More precisely, Low assumed that, $A = A(\xi, \theta)$. This can now be inverted to $\xi = \xi(A, \theta)$, which is in the form required for the present calculation.
From equation (41) since $r = \xi(A, \theta)\Phi(t)$, so we find

\be\begin{split} & r_t = \xi\Phi_t,~~r_{tt} = \xi\Phi_{tt};~~~~~r_\theta = \xi\Phi_\theta,~~r_{\theta\theta} = \xi\Phi_{\theta\theta};~~~~~r_A = \xi\Phi_A~~r_{AA} = \xi\Phi_{AA}.\end{split}\ee
Now applying equations (42) in equation (33) we obtain,

\be {D_A\xi_\theta\over \Phi^2\xi^2\xi_A}\left(\xi\Phi_{tt} + {G M\over\xi^2\Phi^2}\right)= -\left\{E\left({D_A\over \Phi^3 \xi^2 \xi_A}\right)^\gamma\right\}_\theta.\ee
Applying equation (42) again in equation (40), we obtain

\be \begin{split}&\left[\left({1\over \Phi\xi_A}\right)_A  + {\xi_\theta\over \Phi\xi^2}\left({\xi_\theta\over \xi_A}\right)_A + {\xi_\theta\over \Phi\xi^2}\cot{\theta} + C_A\left({C_A\over \Phi\xi_A}\right)_A + \Phi\xi_A \left\{E\left({D_A\over \Phi^3\xi^2\xi_A}\right)^\gamma\right\}_A\right]\Phi\xi_\theta\\&
 \hspace{3.0 in}= 4\pi \Phi^4 \xi^2 \xi_A^2 \sin^2{\theta}\left\{E\left({D_A\over \Phi^3\xi^2\xi_A}\right)^\gamma\right\}_\theta.\end{split}\ee
Since $\Phi$ is a function of time only, so the above equation can be written as,

\be \begin{split}&\left[\left({1\over \xi_A}\right)_A  + {\xi_\theta\over \xi^2}\left({\xi_\theta\over \xi_A}\right)_A + {\xi_\theta\over \xi^2}\cot{\theta} + C_A\left({C_A\over \xi_A}\right)_A\right]\xi_\theta + \xi_A \left\{E\left({D_A\over \xi^2\xi_A}\right)^\gamma\right\}_A\Phi^{2-3\gamma}\xi_\theta\\&
 \hspace{3.0 in}= 4\pi \xi^2 \xi_A^2 \sin^2{\theta}\left\{E\left({D_A\over\xi^2\xi_A}\right)^\gamma\right\}_\theta  \Phi^{4 - 3\gamma}.\end{split}\ee

\section{Conclusion:}

In summary, in order to find self-similar solutions of magnetohydrodynamics (MHD) in the presence of gravity due to a central point mass, we have started with seven equations viz. (8) to (14) and ultimately completely integrated those equations, and using $A, \theta$ and $t$ as independent variables, all other variables are expressed in their terms as follows. $D$ and $C$ are given by equations (18) and (19), $\Phi(t)$ is obtained from equations (21) and (22), $r$ is defined by equation (1.6) and $E$, by equation (17), where $F(\theta)$ is arbitrary. $v, B, p$ and $p$ are obtained from equations (5), (6)(8) and (9), respectively. So now one can trivially make changes of variables to express $A, B, v, p$ and $p$ in terms of $r, \theta, t$. We have also found that the MHD equations give physically permissible solutions only when $\gamma = {4\over 3}$ polytrope and also for self-similar solutions, pressure becomes a function of $\theta$ and $t$ only. Self-similar solutions of the MHD equations have got wide applications in stellar structure. Low (1982) in his paper, though did not obtain a complete set of solutions, yet discussed the probable applications of the MHD equations in detail, in 130 connection with the coronal transient.\\
\noindent
Acknowledgement: Thanks are due to Professor B. N. BASU of the Department of Applied Mathematics, University College of Science, Calcutta, for some useful discussions.

\end{document}